\documentclass[prmaterials,letterpaper,amsmath,amssymb,superscriptaddress,reprint]{revtex4-1}
\usepackage{graphicx}
\usepackage{microtype}
\usepackage[utf8]{inputenc} 
\usepackage{float}

\begin{document}
\title{Improving thermal stability of MnN/CoFeB exchange bias systems by optimizing the Ta buffer layer}

\author{M.\,Dunz}
 \email{mdunz@physik.uni-bielefeld.de}
\affiliation{ 
Center for Spinelectronic Materials and Devices, Department of Physics, Bielefeld University, D-33501 Bielefeld, Germany 
}%
\author{M.\,Meinert}%
\affiliation{ 
Department of Electrical Engineering and Information Technology, Technical University of Darmstadt, Merckstraße 25, D-64283 Darmstadt, Germany
}
\date{\today}

\begin{abstract}
We investigated the influence of the Ta buffer layer on the thermal stability of polycrystalline Ta/~MnN/~CoFeB exchange bias systems, showing high exchange bias of about 1800\,Oe at room temperature. The thermal stability of those trilayer systems is limited by nitrogen diffusion that occurs during annealing processes. Most of the nitrogen diffuses into the Ta buffer layer, which is necessary for good crystal growth of MnN and thus a crucial component of the exchange bias system. In order to improve the thermal stability, we prepared exchange bias stacks where we varied the Ta thickness to look for an optimum value that guarantees stable and high exchange over a broad temperature range. Our findings show that thin layers of $2-5$\,nm Ta indeed support stable exchange bias up to annealing temperatures of more than  $550^{\circ}$\,C. Furthermore, we found that the introduction of a TaN$_{\text{x}}$ layer between MnN and Ta, acting as a barrier, can prevent nitrogen diffusion. However, our results show that those measures, even though being beneficial in terms of thermal stability, often lead to decreased crystallinity and thus lower the exchange bias. 
\end{abstract}

\maketitle


\section{\label{sec:level1}Introduction}
The exchange bias effect\,\citep{Meiklejohn, Schuller, Berkowitz, Kiwi, Grady} is used in spin electronic devices to pin a ferromagnetic electrode to an antiferromagnetic layer. This is crucial in GMR or TMR stacks to allow for distinct stable resistance states\,\citep{Chappert}. For several years, the search for new antiferromagnetic materials for exchange bias has been going on in order to find rare-earth free alternatives for commonly used MnIr\,\citep{Funke} or MnPt\,\citep{Saito, Glaister}. For device integration, the antiferromagnet should be easy to prepare, generate exchange bias fields that are clearly higher than corresponding coercive fields and be thermally stable at typical device operation temperatures. As we recently reported\,\citep{Meinert2015, Zilske, Dunz1, Mewes}, antiferromagnetic MnN is a very promising candidate.
MnN crystallizes in the $\theta-$phase of the Mn$-$N phase diagram\,\citep{Gokcen}, a tetragonal variant of the NaCl structure with $a = b = 4.256$\,\AA\, and $c = 4.189$\,\AA\, at room temperature\,\citep{Suzuki2000}. The exact lattice constants depend on the nitrogen content in the lattice. With increasing nitrogen content, increasing lattice constants are observed\,\citep{Suzuki2000, Leineweber}. In optimized stacks, MnN generates exchange bias of up to 1800 Oe at room temperature with an effective interfacial exchange energy of $J_\mathrm{eff} = 0.41$\,mJ/m$^2$\,\citep{Meinert2015}, satisfying above mentioned requirements for integration into spin electronic devices. The Néel temperature of MnN is around $660$\,K\,\citep{Tabuchi} and Ta/~MnN/~CoFe systems show a median blocking temperature of $160\,^{\circ}$C\,\citep{Meinert2015}. However, nitrogen diffusion at high temperatures, respectively long annealing times, limits the thermal stability of the system\,\citep{Sinclair}. In the course of our previous investigations\,\citep{Meinert2015} we already found that preparing MnN with a higher nitrogen concentration can slightly increase the thermal stability but at the same time lowers the exchange bias. Similar results are obtained when doping MnN with elements that have stronger bonds with nitrogen\,\citep{Dunz2}.

Recent polarized neutron reflectometry\,\citep{Quarterman} and Auger electron depth profile\,\citep{Dunz1} studies of Ta 10\,nm/ MnN 30\,nm/ CoFeB 1.6\,nm stacks revealed the crucial role of the Ta buffer layer on which MnN is grown for better crystallinity. By investigating the samples directly after preparation as well as after annealing, it was found that the nitrogen diffuses from MnN into the Ta buffer layer during the annealing process. After annealing at high temperatures of $T_{\text{A}}>500^{\circ}$\,C, the Ta layer is fully saturated with nitrogen. It was proposed that, in terms of better thermal stability, using a thinner Ta buffer layer could be beneficial\,\citep{Quarterman}. The thinner the Ta, the less nitrogen needs to diffuse from the MnN to the Ta layer for the latter to be saturated. In the present article, we thus study the influence of the Ta buffer layer on the thermal stability of Ta/~MnN/~CoFeB exchange bias systems. We perform annealing series for samples with varying Ta thickness and furthermore investigate the effects of introducing an additional TaN$_{\text{x}}$ layer between MnN and Ta. This extra layer could act as a barrier preventing nitrogen diffusion from MnN to Ta. 


\section{\label{sec:level1}Experimental}
We prepared two different sets of samples. In the first set, the Ta buffer layer thickness was varied between $t_{\text{Ta}}=1-15$\,nm, resulting in a sample setup of  Ta $t_{\text{Ta}}$/ MnN 30\,nm/ Co$_{40}$Fe$_{40}$B$_{20}$ 1.6\,nm/ Ta 0.5\,nm/ Ta$_2$O$_5$ 2\,nm. For the second set of samples, an additional TaN$_{\text{x}}$ layer was introduced while the Ta thickness was kept constant at 3 or 10\,nm: Ta 3 or  10\,nm/ TaN$_{\text{x}}$ $t_{\text{TaN}_{\text{x}}}$/ MnN 30\,nm/ Co$_{40}$Fe$_{40}$B$_{20}$ 1.6\,nm/ Ta 0.5\,nm/ Ta$_2$O$_5$ 2\,nm. The thickness of TaN$_{\text{x}}$ was varied from $0.5-3$\,nm.
Both sample sets were prepared on thermally oxidized SiO$_{\text{x}}$ substrates via magnetron (co-)sputtering at room temperature. We followed exactly the same preparation procedure as described in our previous report\,\citep{Meinert2015}. The base pressure of the sputtering system was around $5 \times 10^{-9}$\,mbar prior to the deposition runs. The MnN films were reactively sputtered from an elemental Mn target with a gas ratio of  $50\%$ Ar to $50\%$ N$_2$ at a working pressure of $p_{\text{w}}=2.3\cdot10^{-3}$\,mbar. The typical deposition rate of MnN was 0.1\,nm/s at a source power of 50\,W. The intermediate TaN$_{\text{x}}$ layer was also prepared via reactive sputtering from an elemental Ta target. A gas ratio of  $78\%$ Ar to $22\%$ N$_2$ was used, yielding the best crystallographic results throughout a partial pressure series from $14\%$ to $22\%$ nitrogen. Subsequent post-annealing for 15 min and field cooling in a magnetic field of $H_{\text{fc}}=6.5$\,kOe parallel to the film plane was performed in a vacuum furnace with pressure below $5\cdot 10^{-6}$\,mbar to activate exchange bias. Magnetic characterization of the samples was performed using the longitudinal magneto-optical Kerr effect at room temperature. For annealing series, samples were successively annealed and measurements were taken in between the single annealing steps. Structural analysis was performed via X-ray diffraction with a Philips X'Pert Pro MPD, which is equipped with a Cu source and Bragg-Brentano optics.


\section{\label{sec:level1}Results and Discussion}

\subsection{Ta thickness variation}

\begin{figure}
\centering
\includegraphics[scale=0.56]{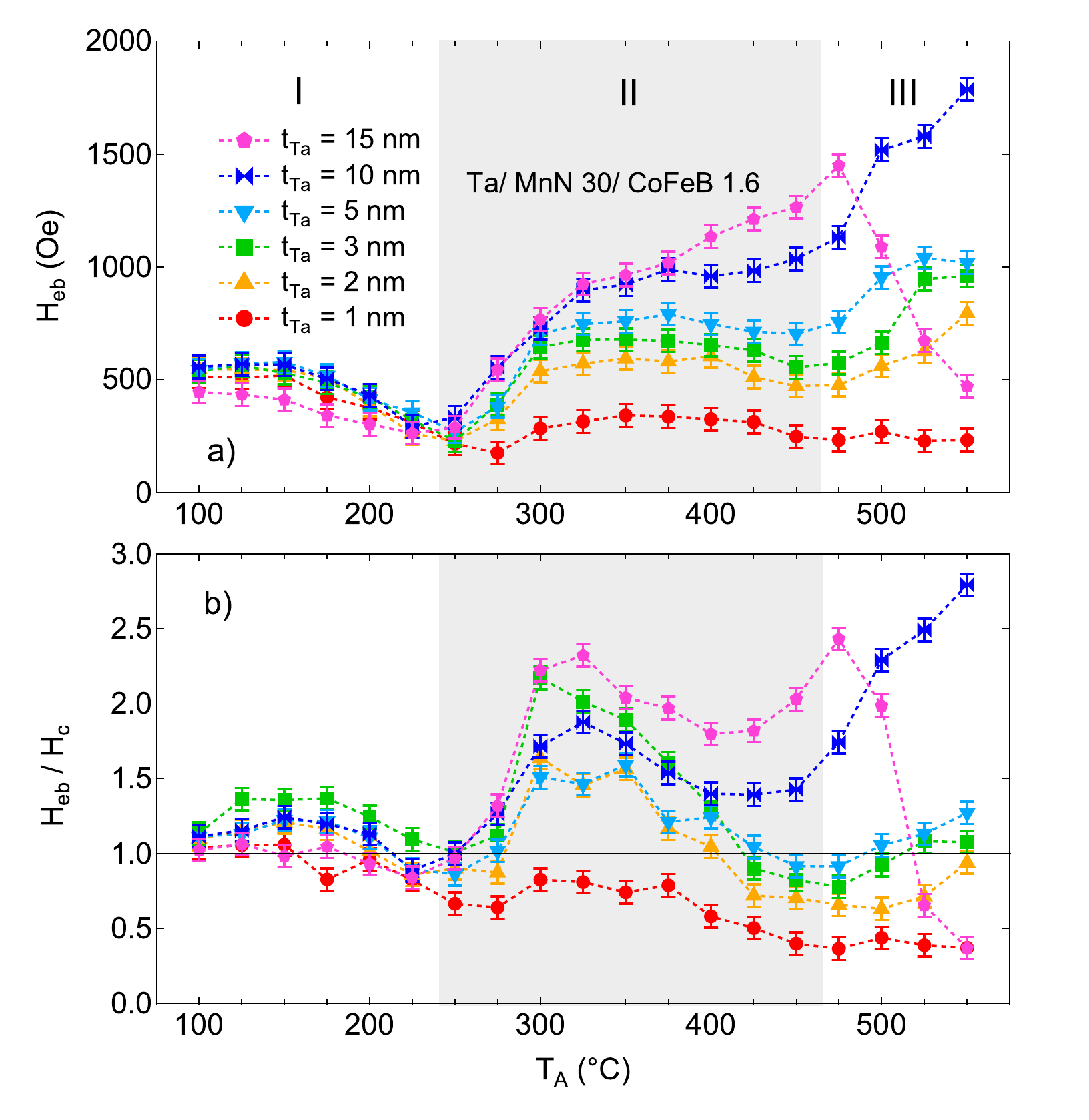}
\caption{\label{Fig1}a) Exchange bias $H_{\text{eb}}$ and b) ratio $H_{\text{eb}}/H_{\text{c}}$ as a function of annealing temperature $T_{\text{A}}$ for Ta/~MnN/~CoFeB exchange bias systems with varying Ta thickness. Measurements were taken after each annealing step at room temperature. I, II and III mark three distinct annealing temperature regimes.}
\end{figure}

To investigate how the Ta buffer layer thickness influences the thermal stability of Ta/~MnN/~CoFeB exchange bias systems, annealing series with temperatures ranging from $T_{\text{A}}= 100^{\circ}$\,C to $550^{\circ}$\,C were performed. Fig.~\ref{Fig1}a) displays the dependence of exchange bias on the annealing temperature for samples with Ta thicknesses varying between 1\,nm and 15\,nm. As we chose CoFeB as a ferromagnetic layer instead of CoFe, the total exchange bias values are somewhat smaller and the optimized sample with the commonly used 10\,nm Ta slightly deviates from the results observed in our previous studies\,\citep{Meinert2015, Dunz1, Dunz2}. Notably, independent of the Ta thickness, all samples follow a similar behavior at low annealing temperatures. Starting from exchange bias values of 500\,Oe, a minimum is reached around $T_{\text{A}}= 250^{\circ}$\,C. We identify this temperature range as the first of three regimes, indicated by background coloring in the graph. At higher annealing temperatures, the exchange bias rises and reaches a plateau. In this second regime, the Ta thickness obviously has a strong influence: the thicker the Ta layer, the higher the exchange bias. Thicker Ta layers yield stronger diffusion from MnN into the Ta during annealing. However, up until temperatures around $T_{\text{A}}=450^{\circ}$\,C,  this process does not lower the exchange bias, as can be seen from the results for the sample with 15\,nm Ta. Only when reaching the third regime with high annealing temperatures of $T_{\text{A}}>450^{\circ}$\,C, nitrogen diffusion becomes a crucial factor. The sample with the thickest Ta buffer layer loses its exchange bias whereas the other samples do not show decreasing values yet. Unfortunately, the furnace used for this annealing series cannot go to higher temperatures. We speculate that next the sample with 10\,nm Ta loses exchange bias after reaching a maximum around $T_{\text{A}}=550^{\circ}$\,C. This hypothesis is supported by the fact that all samples, except for the thinnest one, show a very similar dependence on the annealing temperaure: After reaching the plateau in regime II, a local minimum can be observed before the exchange bias rises again in regime III. This minimum and the corresponding rise are located at higher annealing temperatures for samples with thinner Ta layers. Possibly, those samples exhibit a maximum of exchange bias at temperatures $T_{\text{A}}>550^{\circ}$\,C, followed by a collapse, that might also be shifted to higher temperatures. The minimum of exchange bias marking the transition from temperature range I to II has also been observed in all our previous studies on MnN exchange bias systems\,\citep{Meinert2015, Dunz1, Dunz2}. At this point we are not sure what causes this behavior, but it is likely that a magnetic ordering transition is happening at annealing temperatures around $T_{\text{A}}= 250^{\circ}$\,C. The fact that all samples show similar values of exchange bias in the first regime could hint at the formation of a spin glass structure in the MnN\,\citep{Kouvel, Ali}. We suppose that the AF-I antiferromagnetic order is developed upon transition into regime II. 
Fig.~\ref{Fig1}b) shows the ratio of exchange bias and corresponding coercive field $H_{\text{eb}}/H_{\text{c}}$ as a function of annealing temperature. $H_{\text{eb}}/H_{\text{c}} > 1$ needs to be fullfilled for any exchange bias system to be integratable into spintronic devices. Here, this is reached for all samples with Ta thicknesses higher than 1\,nm. However, for $T_{\text{A}}>400^{\circ}$\,C the ratio of the 2\,nm Ta sample decreases to values smaller than one, followed by samples with 3 and 5\,nm Ta layers.

\begin{figure}
\centering
\includegraphics[scale=0.55]{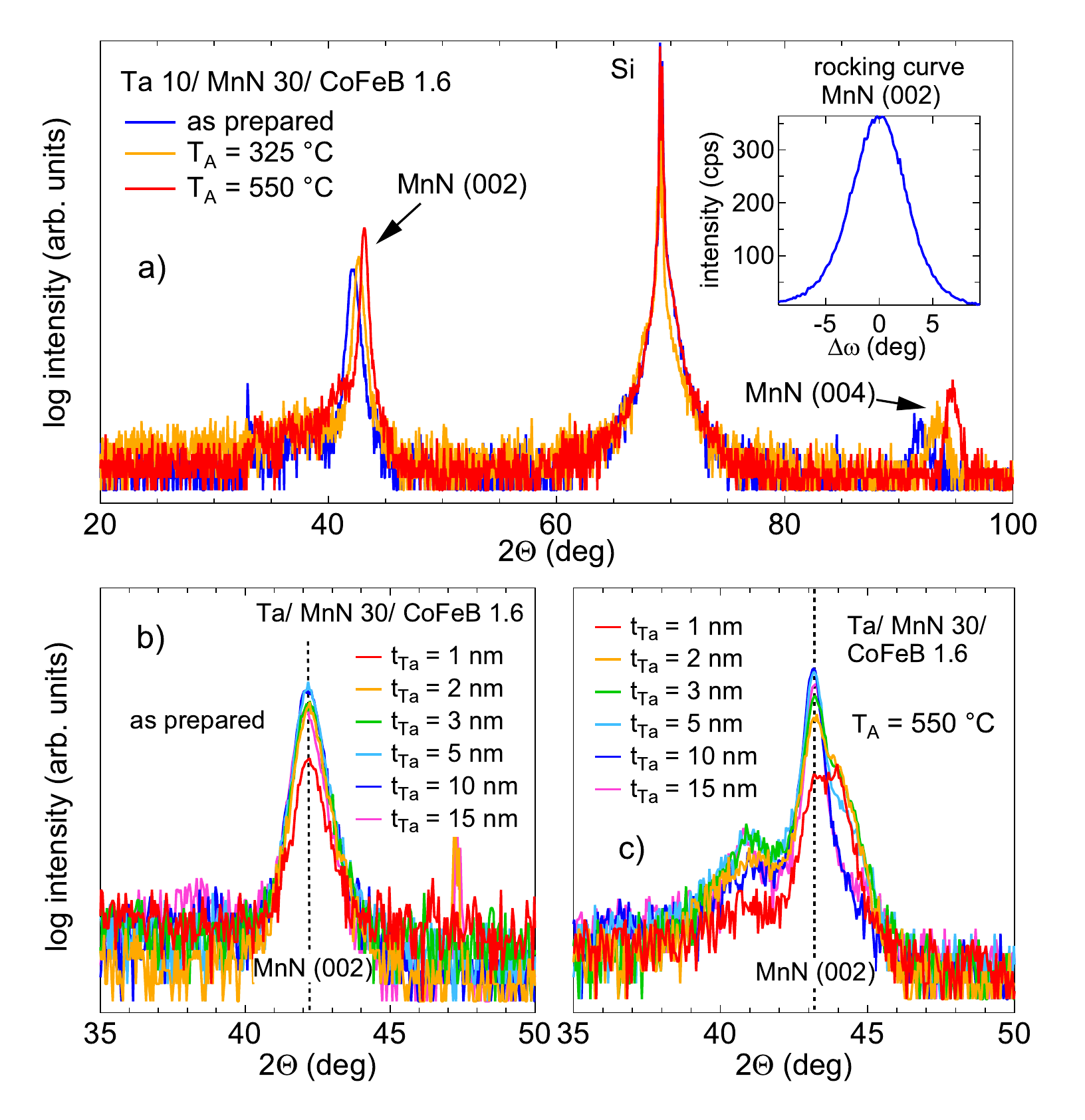}
\caption{\label{Fig2}a) X-ray diffraction pattern of a Ta/~MnN/~CoFeB sample with a 10\,nm Ta buffer layer as prepared and after annealing at $325^{\circ}$\,C, and $550^{\circ}$\,C. The inset shows the rocking curve of MnN's (002) peak after preparation. b)-c) Diffraction scans around the angular range of the (002) peak of MnN for different Ta thicknesses detected b) directly after preparation and c) after annealing at $550^{\circ}$\,C.}
\end{figure}

Next, the influence of the Ta thickness on the crystallographic properties of MnN was investigated via X-ray diffraction. Fig.~\ref{Fig2}a) shows diffraction patterns of the Ta/~MnN/~CoFeB system with a 10\,nm Ta buffer layer directly after preparation, after annealing at $T_{\text{A}}= 325^{\circ}$\,C (regime II), and after annealing at the highest possible temperature of $T_{\text{A}}= 550^{\circ}$\,C (regime III). The measurements reveal a polycrystalline, columnar growth of MnN in (001) direction. The lattice constant of MnN directly after preparation is $c = 4.282$\,\AA, which is slightly larger than the bulk values reported in literature\,\citep{Suzuki2000}. The inset shows the corresponding rocking curve of MnN's (002) peak, which yields a FWHM of $5.92^{\circ}$ for the as prepared sample. This suggests a rather uniform growth of the crystallites in the $c-$direction. After annealing, the MnN peaks are slightly shifted to higher angles, indicating a smaller lattice parameter of $c = 4.189$\,\AA\, after annealing at $T_{\text{A}}= 550^{\circ}$\,C. This can be attributed to the generation of nitrogen vacancies as interdiffusion occurs during the annealing process, leading to a decrease of MnN's lattice constants\,\citep{Meinert2015}. In Figs.~\ref{Fig2}b) and c), diffraction scans around the angular range of MnN's (002) peak are shown for all Ta thicknesses. Directly after preparation (Fig.~\ref{Fig2}b)), MnN yields good crystallinity in all samples, except for the one with 1\,nm Ta. The reduced crystallinity is directly connected to the low exchange bias values observed for this sample. After annealing at $T_{\text{A}}= 550^{\circ}$\,C (Fig.~\ref{Fig2}c)), significant changes to the crystal structure are visible. The MnN's (002) peak narrows, indicating vertical crystallite growth or relaxation of microstrain. In our previous study\,\citep{Meinert2015}, we concluded that the crystallites are columnar and their height corresponds to the film thickness; thus, the narrowing here can be assigned to a relaxation of microstrain. However, in all samples with $t_{\text{Ta}}<10$\,nm an additional peak arises in the right shoulder of the MnN (002) peak. The fact that this peak is not observable for all samples suggests that it is connected to a structural transition that only occurs if the Ta layer is thin enough, i.e. enough nitrogen is left in the MnN layer. An additional peak emerges around $41^{\circ}$ in all samples. Possibly, those peaks result from an emerging phase transition of the MnN to a less nitrogen rich composition. Alternatively, they could be related to the formation of a crystalline TaN$_{\text{x}}$ interlayer between Ta and MnN. We will discuss the origin of those unidentified peaks in more detail below. 

\subsection{TaN$_{\text{x}}$ interlayer}
\begin{figure}
\centering
\includegraphics[scale=0.55]{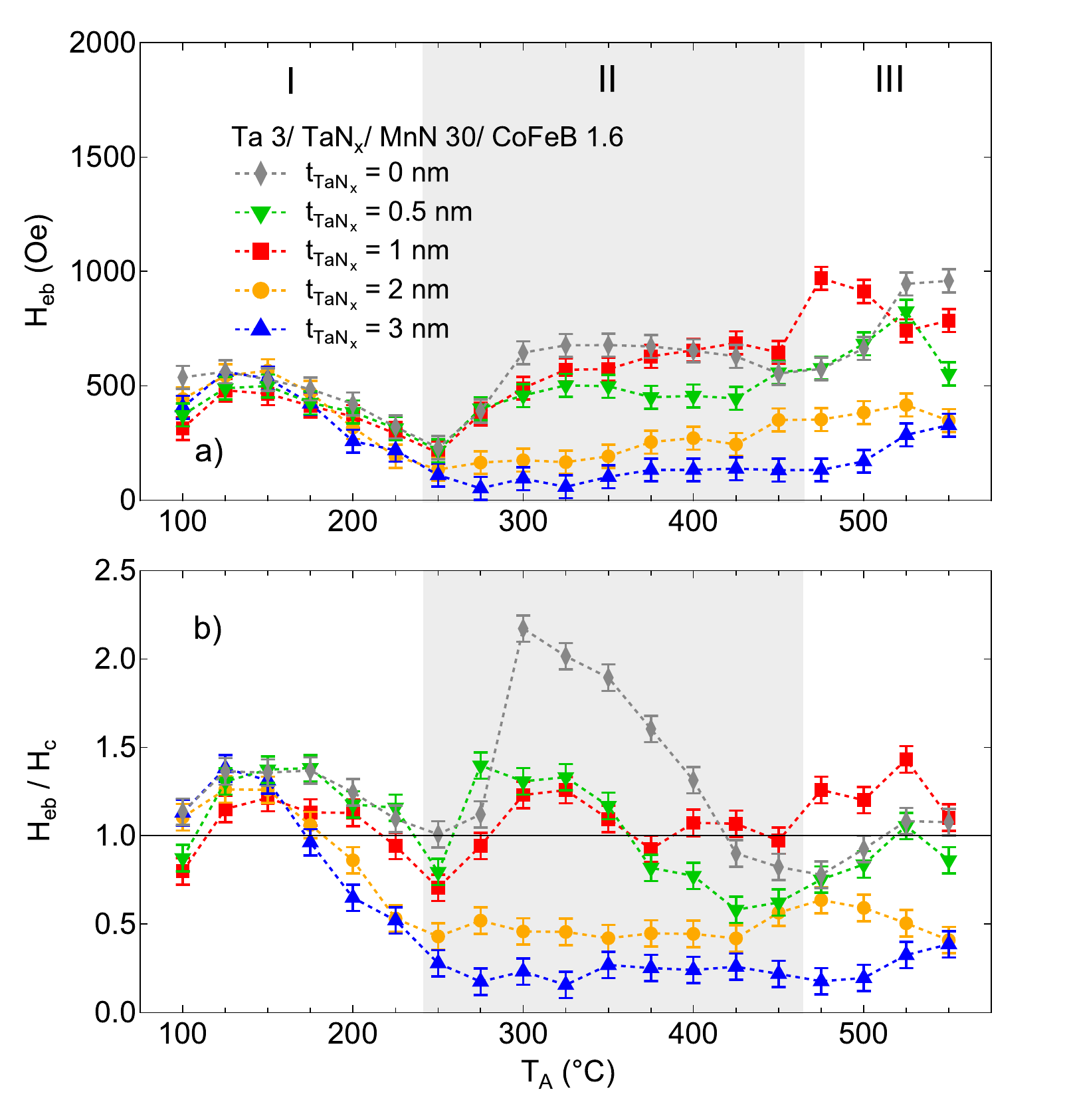}
\caption{\label{Fig3}a) Exchange bias $H_{\text{eb}}$ and b) ratio $H_{\text{eb}}/H_{\text{c}}$ as a function of annealing temperature $T_{\text{A}}$ for Ta/~MnN/~CoFeB exchange bias samples with a 3\,nm buffer layer of Ta and an additional TaN$_{\text{x}}$ interlayer of varying thickness. Measurements were taken after each annealing step at room temperature. I, II and III mark the three annealing temperature regimes.}
\end{figure}

\begin{figure}
\centering
\includegraphics[scale=0.55]{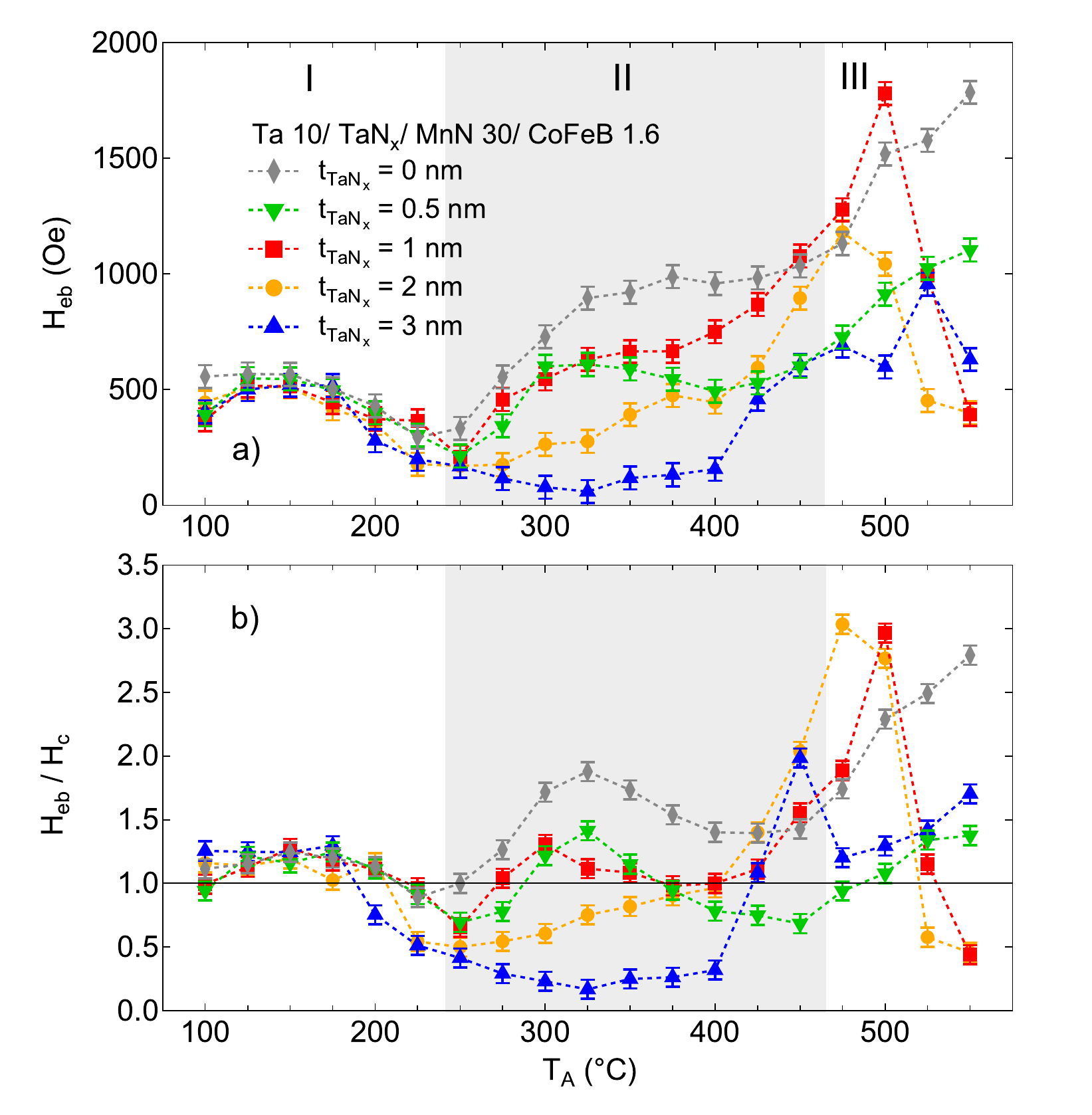}
\caption{\label{Fig4}a) Exchange bias and b) ratio $H_{\text{eb}}/H_{\text{c}}$ as a function of annealing temperature $T_{\text{A}}$ for Ta/~MnN/~CoFeB exchange bias samples with a 10\,nm buffer layer of Ta and an additional TaN$_{\text{x}}$ interlayer of varying thickness. Measurements were taken after each annealing step at room temperature. I, II and III mark the three annealing temperature regimes.}
\end{figure}

To further understand the role of the buffer layer in the diffusion processes occuring in the Ta/~MnN/~CoFeB exchange bias system, we prepared a second set of samples with an additional TaN$_{\text{x}}$ interlayer. While the Ta thickness is kept constant at either 3\,nm or 10\,nm, the TaN$_{\text{x}}$ thickness is varied between 0.5\,nm and 3\,nm. Fig.~\ref{Fig3}a) shows the dependence of exchange bias on the annealing temperature for samples with $t_{\text{Ta}}=3$\,nm and varying TaN$_{\text{x}}$ thickness. Again, the results can be separated into three different temperature regimes, indicated in the graph. For low annealing temperatures, all samples show exchange bias values around $500$\,Oe before reaching a minimum around $T_{\text{A}}= 250^{\circ}$\,C. This is similar to the behavior observed for the Ta thickness series discussed above and yet another hint that a magnetic ordering transition is happening between temperature regimes I and II. At higher annealing temperatures, the influence of the TaN$_{\text{x}}$ interlayer gets more significant. Whereas the samples with 0.5\,nm and 1\,nm TaN$_{\text{x}}$ show increasing exchange bias values and reach a plateau, a thicker interlayer of TaN$_{\text{x}}$ causes the absence of a local maximum in regime II. The results observed for even higher annealing temperatures need to be interpreted very carefully. Diffusion  processes and structural transitions are happening simultaneously, making it hard to disentagle their effect on the exchange bias. All samples show increasing exchange bias for $T_{\text{A}}> 450^{\circ}$\,C. However, the two samples with thinner TaN$_{\text{x}}$ exhibit a distinct maximum, whereas the samples with thicker TaN$_{\text{x}}$ only show a small increase. As can be seen in Fig.~\ref{Fig3}b), the requirement for the ratio $H_{\text{eb}}/H_{\text{c}}$ to be larger than one is not fullfilled for any sample with a TaN$_{\text{x}}$ interlayer throughout a broad range of the investigated temperature window.

The results of the annealing series for the samples with $t_{\text{Ta}}=10$\,nm and varying TaN$_{\text{x}}$ thickness are displayed in Fig.~\ref{Fig4}. The development of exchange bias (Fig.~\ref{Fig4}a)) resembles the one observed for $t_{\text{Ta}}=3$\,nm. Exchange bias values lie around $500$\,Oe for all samples in the first temperature regime. This is followed by strongly varying behavior for the different TaN$_{\text{x}}$ thicknesses in regime II. The two samples with the thickest TaN$_{\text{x}}$ interlayers exhibit low exchange bias without a local maximum in the second regime while the samples with thinner TaN$_{\text{x}}$ reach the typical plateau. At higher annealing temperatures, all samples show increasing exchange bias. The behavior in the third temperature regime seems to be even more complicated here than it was for $t_{\text{Ta}}=3$\,nm. Due to the thicker Ta buffer layer, i.e. larger nitrogen drain, diffusion becomes especially important. The samples with 1, 2 and 3\,nm TaN$_{\text{x}}$ exhibit a maximum around $T_{\text{A}}> 500^{\circ}$\,C and then lose their exchange bias, whereas the one with only 0.5\,nm TaN$_{\text{x}}$ shows steadily increasing values. This observation stands in constrast to our hypothesis that thicker TaN$_{\text{x}}$ layers yield better thermal stability. Possibly, there are phase transistions happening in the MnN layer that support stable exchange bias in the sample with 0.5\,nm TaN$_{\text{x}}$. Overall, the exchange bias values are slightly higher in the case of $t_{\text{Ta}}=10$\,nm than for $t_{\text{Ta}}=3$\,nm. This can be attributed to the better crystallinity of MnN when thicker Ta layers are used, which goes in line with the results of the Ta thickness series discussed above. Nevertheless, mostly low ratios $H_{\text{eb}}/H_{\text{c}}$ are observed (Fig.~\ref{Fig4}b)). Values of one are only exceeded in some parts of the temperature range, depending on the TaN$_{\text{x}}$ thickness. However, these results show that a specific TaN$_{\text{x}}$ thickness can be used to tune the thermal stability at a given temperature. For example, using 2\,nm TaN$_{\text{x}}$, especially high ratios are observed for $T_{\text{A}}=450-500^{\circ}$\,C.

\begin{figure}
\centering
\includegraphics[scale=0.55]{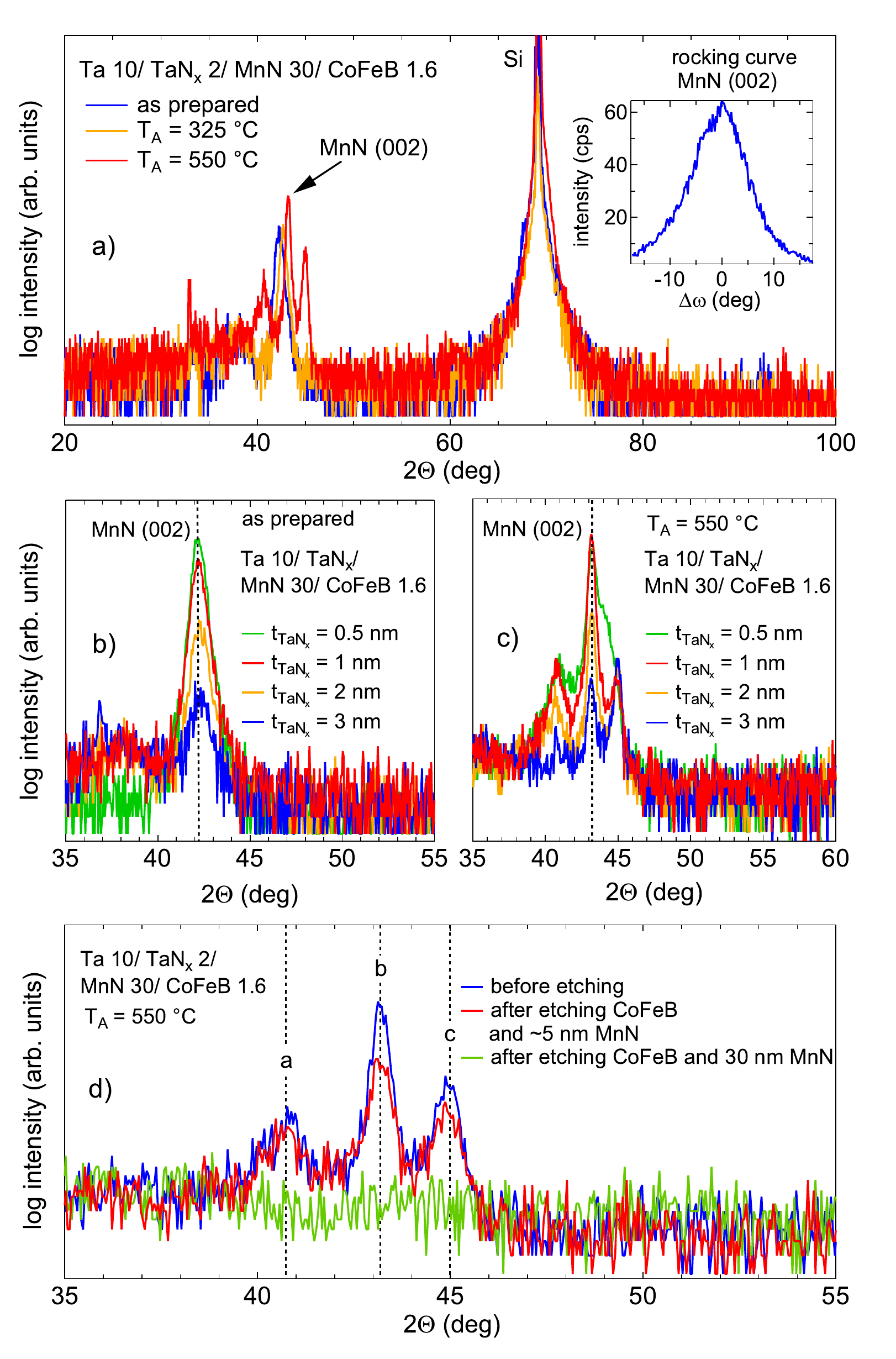}
\caption{\label{Fig5}a) X-ray diffraction pattern of a Ta/~MnN/~CoFeB sample with 10\,nm Ta and a 2\,nm TaN$_{\text{x}}$ interlayer as prepared and after annealing at $325^{\circ}$\,C, and $550^{\circ}$\,C. The inset shows the rocking curve of MnN's (002) peak directly after preparation. b)-c) Diffraction scans around the angular range of MnN's (002) peak for different TaN$_{\text{x}}$ thicknesses on 10\,nm Ta detected b) directly after preparation and c) after annealing at $550^{\circ}$\,C. d) Diffraction scan of a Ta/~MnN/~CoFeB sample with 10\,nm Ta and 2\,nm TaN$_{\text{x}}$ after annealing at $550^{\circ}$\,C for the full stack and after etching layers off.}
\end{figure}

To gain a deeper understanding about the influence of diffusion and crystallization processes, especially at high annealing temperatures, X-ray diffraction scans were performed for all samples with $t_{\text{Ta}}=10$\,nm and varying TaN$_{\text{x}}$ thickness. In Fig.~\ref{Fig5}a), the diffraction patterns measured for the sample with $t_{\text{Ta}}=10$\,nm and $t_{\text{TaN}_{\text{x}}}=2$\,nm directly after preparation and after annealing at $T_{\text{A}}= 325^{\circ}$\,C (regime II) and $T_{\text{A}}= 550^{\circ}$\,C (regime III) are shown. The lattice constant of MnN before annealing is $c = 4.276$\,\AA, which agrees well with the value determined for the sample without an additional layer of TaN$_{\text{x}}$. Again, MnN's lattice constant decreases with increasing annealing temperature, yielding $c = 4.185$\,\AA\, after annealing at $T_{\text{A}}= 550^{\circ}$\,C, which also matches the result obtained for the sample without TaN$_{\text{x}}$. This means an interlayer of TaN$_{\text{x}}$ does not change the crystallographic properties of MnN in terms of lattice parameters. Yet, the rocking curve of MnN's (002) peak, displayed in the inset, reveals the main crystallographic transformation that TaN$_{\text{x}}$ induces. With $11.47^{\circ}$, the FWHM of MnN's (002) peak is more than twice as large as in a similar sample without TaN$_{\text{x}}$. Using an interlayer of 3\,nm TaN$_{\text{x}}$, it even increases to $18.6^{\circ}$. The thicker the TaN$_{\text{x}}$ layer is, the more of MnN's crystallites are tilted away from the film normal. MnN grows less uniformly orientated and with a probably smaller lateral grain diameter, resulting in reduced exchange bias values. Remarkably, judging solely from the X-ray diffraction results, TaN$_{\text{x}}$ does not seem to prevent nitrogen diffusion significantly, as the lattice constant of MnN decreases similarly strong after annealing, with or without the layer of 2\,nm TaN$_{\text{x}}$. However, other changes to the structure are observable after annealing at $T_{\text{A}}= 550^{\circ}$\,C. Two additional peaks arise near the MnN's (002) peak. 

In Fig.~\ref{Fig5}b), diffraction scans around the angular range of the MnN (002) peak for all TaN$_{\text{x}}$ thicknesses directly after preparation are displayed. It is obvious that with increasing TaN$_{\text{x}}$ thickness the MnN's crystallinity decreases. This is in line with the results of the rocking curves and the observation of very low exchange bias for the samples with $t_{\text{TaN}_{\text{x}}}=2$\,nm and 3\,nm. After annealing at $T_{\text{A}}= 550^{\circ}$\,C, the MnN peak narrows and additional peaks arise for all TaN$_{\text{x}}$ thicknesses (Fig.~\ref{Fig5}c)). Notably, the MnN (002) peak's intensity decreases with increasing TaN$_{\text{x}}$ thickness, whereas the peak at $45^{\circ}$ is nearly independent of the TaN$_{\text{x}}$ thickness. Only for the sample with 0.5\,nm TaN$_{\text{x}}$ it is located at smaller angles and therefore lies in the shoulder of the MnN peak. This curve looks remarkably similar to the diffraction results obtained for the Ta thickness series in Fig.~\ref{Fig2}c). However, there, the peak in the MnN (002) peak's shoulder was only observed for $t_{\text{Ta}}<10$\,nm. Even a thin interlayer of 0.5\,nm TaN$_{\text{x}}$ thus significantly changes the crystallographic properties. To clarify whether the additional peaks that are observed after high-temperature annealing originate from the Ta(N$_{\text{x}}$), the MnN or the ferromagnetic layer, we performed further diffraction measurements after subsequently etching off single layers with Ar ion bombardement. The results are displayed in Fig.~\ref{Fig5}d) for the sample with $t_{\text{Ta}}=10$\,nm and $t_{\text{TaN}_{\text{x}}}=2$\,nm. First, the capping layer, the CoFeB layer and about 5\,nm MnN were etched off. All three peaks are still detectable. Thus, we can conclude that they do not result from the formation of a nitride in the ferromagnetic layer. However, the $\theta-$MnN's (002) peak (b) loses a significant amount of intensity whereas peak (a)'s intensity only decreases weakly and peak (c) is nearly not influenced at all. These observations suggest that peaks (a) and (c) relate to crystallites that are closer to the MnN/ TaN$_{\text{x}}$ interface than to the CoFeB/~MnN interface. Next, the complete MnN layer was etched off, so that Ta and TaN$_{\text{x}}$ are the only layers left on the sample. This leads to the total disappearance of all peaks. Consequently, peak (a) and (c) must be originating from other, nitrogen-poor, Mn$-$N phases that form during annealing. Peak (a) could be identified as the (111) peak of ferrimagnetic $\varepsilon-\text{Mn}_4\text{N}$\,\citep{Chang} or the (002) peak of $\zeta-\text{Mn}_2\text{N}$\,\citep{Meng}. These nitrogen-poor phases might form very close to the TaN$_{\text{x}}$ interface as much of the nitrogen has been lost due to diffusion here. Peak (c) could correspond to the (200) or (006) peak of antiferromagnetic $\eta-\text{Mn}_3\text{N}_2$\,\citep{Yang}. A growth in (100) direction yields the in-plane orientation of the $c-$axis. This would result in a compensated interface between $\eta-\text{Mn}_3\text{N}_2$ and CoFeB as well as magnetic moments aligning perpendicular to the film plane, leading to a reduction of exchange bias\,\citep{Leineweber}.

Summarizing the effects of introducing a TaN$_{\text{x}}$ layer, it gets clear that there is no optimum TaN$_{\text{x}}$ thickness that guarantees high exchange bias and increased thermal stability at the same time. Thick TaN$_{\text{x}}$ layers reduce the crystalline quality of MnN and thus also the exchange bias. In fact, the TaN$_{\text{x}}$ might not only act as barrier for nitrogen diffusion but also induce phase transitions during annealing that are not fully understood at this point. In any case, the detrimental effects of the TaN$_{\text{x}}$ interlayer on the crystallography of the MnN films seem to outweigh possible positive effects of inhibiting of nitrogen diffusion. However, crystallographic quality does not seem to be the only driver of high exchange bias in this complex trilayer system. 


\section{\label{sec:level1}Conclusion}
In summary, we increased the thermal stability of Ta/~MnN/~CoFeB exchange bias systems by reducing nitrogen diffusion from MnN into the Ta buffer layer. Therefor, we studied the influence of the Ta buffer layer's thickness on exchange bias throughout a large annealing temperature range and found that samples with thinner Ta layers yield more stable exchange bias at higher temperatures. However, MnN's crystallinity worsens with decreasing Ta thickness. This results in lower total exchange bias values in samples that are thermally more stable. Furthermore, we introduced a TaN$_{\text{x}}$ diffusion barrier between Ta and MnN. Even though this can promote better thermal stability, it mostly yields low exchange bias values as the crystallinity of MnN is affacted negatively when growing on TaN$_{\text{x}}$. Our results confirm that the Ta buffer layer is a crucial component of the  MnN/~CoFeB exchange bias system. Moreover, they show that the diffusion processes and possible magnetic and structural transitions of MnN that happen during annealing, especially at high temperatures, are more complicated than expected and need further investigation for a comprehensive understanding.

\begin{acknowledgments}
We thank the Ministerium f\"ur Innovation, Wissenschaft und Forschung des Landes Nordrhein-Westfalen (MIWF NRW) for financial support. We further thank G. Reiss for making available laboratory equipment. This work was partially supported by the Deutsche Forschungsgemeinschaft under sign ME 4389/2-1.
\end{acknowledgments}

\end{document}